\documentclass[
reprint,
superscriptaddress,
 amsmath,amssymb, aps,
prb,
floatfix,
showpacs ]{revtex4-1}

\usepackage{graphicx}
\usepackage{color} 
\usepackage{float} 
\usepackage{verbatim}
\usepackage{braket} 
\usepackage[mathscr]{euscript}
\usepackage{silence} 
\usepackage{times}
\usepackage{mathrsfs} 
\usepackage{gensymb}
\usepackage{placeins}
\usepackage{afterpage}
\usepackage{amsthm}
\usepackage{ulem}
\usepackage{longtable}
\usepackage{notes2bib}
\usepackage{bibentry}
\usepackage{xcolor}
\usepackage{algorithm}
\usepackage{lipsum}
\usepackage{algpseudocode} 
\usepackage{hyperref}

 \nobibliography*

\newcommand{\makecopy}[1]{\mathrm{copy}(#1)}            
\newcommand{\R}{\mathbb{R}}
\newcommand{\G}{G}
\renewcommand{\P}{P}
\newcommand{\V}{V}
\newcommand{\B}{B}
\newcommand{\F}{F}
\newcommand{\M}{M}
\newcommand{\N}{N}
\newcommand{\C}{\mathscr{C}}
\newcommand{\Z}{Z}
\newcommand{\Q}{Q}
\newcommand{\I}{I}

\DeclareMathOperator{\stab}{stab}

\newtheorem{theorem}{Theorem}[section]

\newtheorem{definition}[theorem]{Definition}

\let\vec\mathbf

\begin{document}

\preprint{APS/123-QED}

\title{A general algorithm for calculating irreducible Brillouin zones}

\author{Jeremy J. Jorgensen} \affiliation{Department of Physics and Astronomy, Brigham Young University, Provo, Utah, 84602, USA}
\author{John E. Christensen} \affiliation{Department of Physics and Astronomy, Brigham Young University, Provo, Utah, 84602, USA}
\author{Tyler J. Jarvis} \affiliation{Department of Mathematics,
  Brigham Young University, Provo, Utah, 84602, USA}
\author{Gus L. W. Hart} \affiliation{Department of Physics and Astronomy, Brigham Young University, Provo, Utah, 84602, USA}

\break


\date{\today}
\begin{abstract}
Calculations of properties of materials require performing numerical integrals over the Brillouin zone (BZ). Integration points in density functional theory codes are uniformly spread over the BZ (despite integration error being concentrated in small regions of the BZ) and preserve symmetry to improve computational efficiency. Integration points over an irreducible Brillouin zone (IBZ), a rotationally distinct region of the BZ, do not have to preserve crystal symmetry for greater efficiency. This freedom allows the use of adaptive meshes with higher concentrations of points at locations of large error, resulting in improved algorithmic efficiency. We have created an algorithm for constructing an IBZ of any crystal structure in 2D and 3D. The algorithm uses convex hull and half-space representations for the BZ and IBZ to make many aspects of construction and symmetry reduction of the BZ trivial. The algorithm is simple, general, and available as open-source software.
\end{abstract}

\maketitle

\section{Motivation}

\begin{figure}[b]
  \centering
  \includegraphics[width=1.5in]{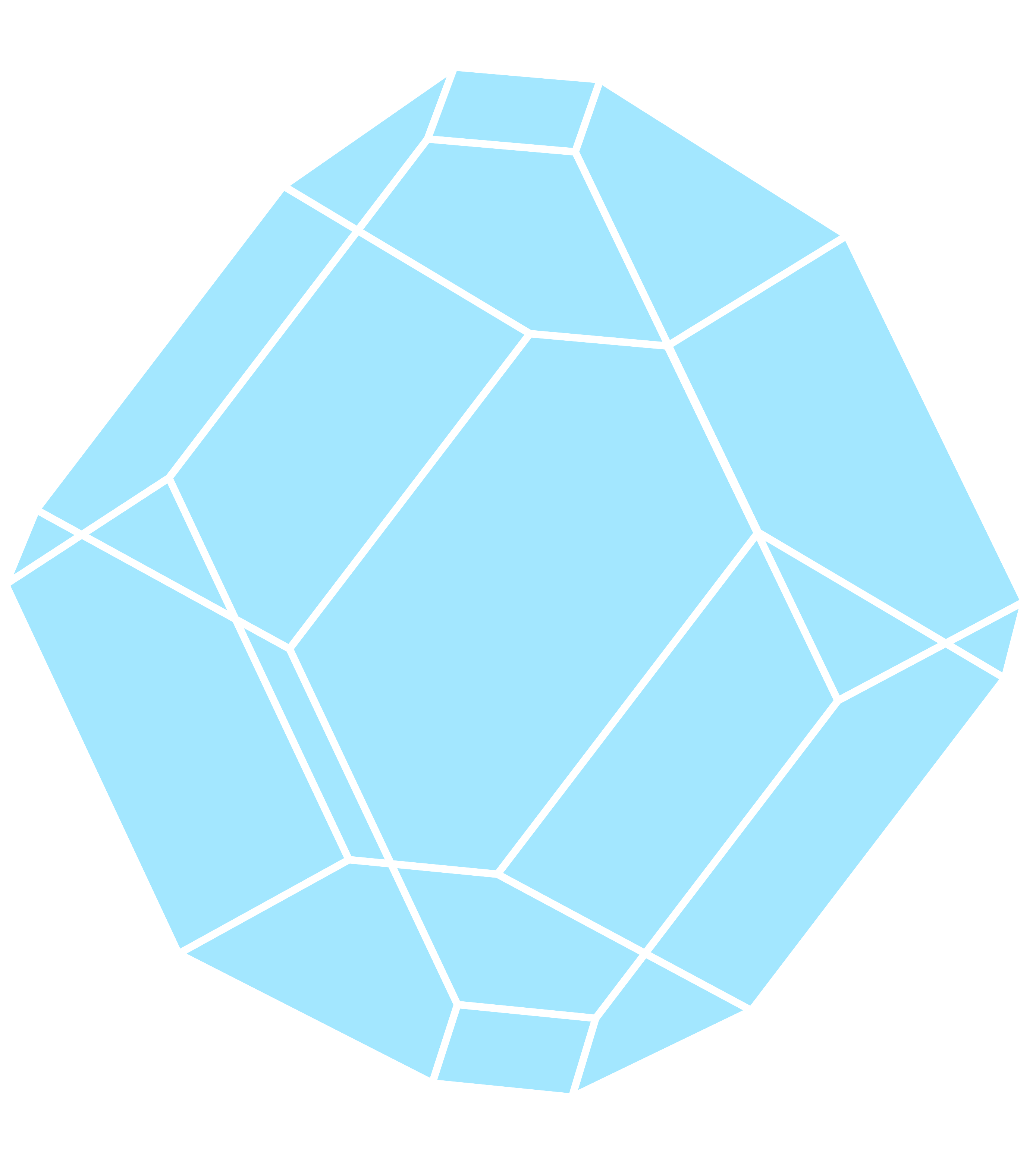}
  \caption{The Brillouin zone for a body-centered tetragonal lattice. The Brillouin zone is the integration domain of integrals that give properties of materials. It is also a convex polygon that often has a complicated shape. This plot, and many others in this article, were created with \href{https://github.com/jerjorg/SymmetryReduceBZ.jl}{\textsc{SymmetryReduceBZ}}.}
  \label{fig:bz}
\end{figure}

\begin{figure}[h]
  \centering
  \includegraphics[width=2in]{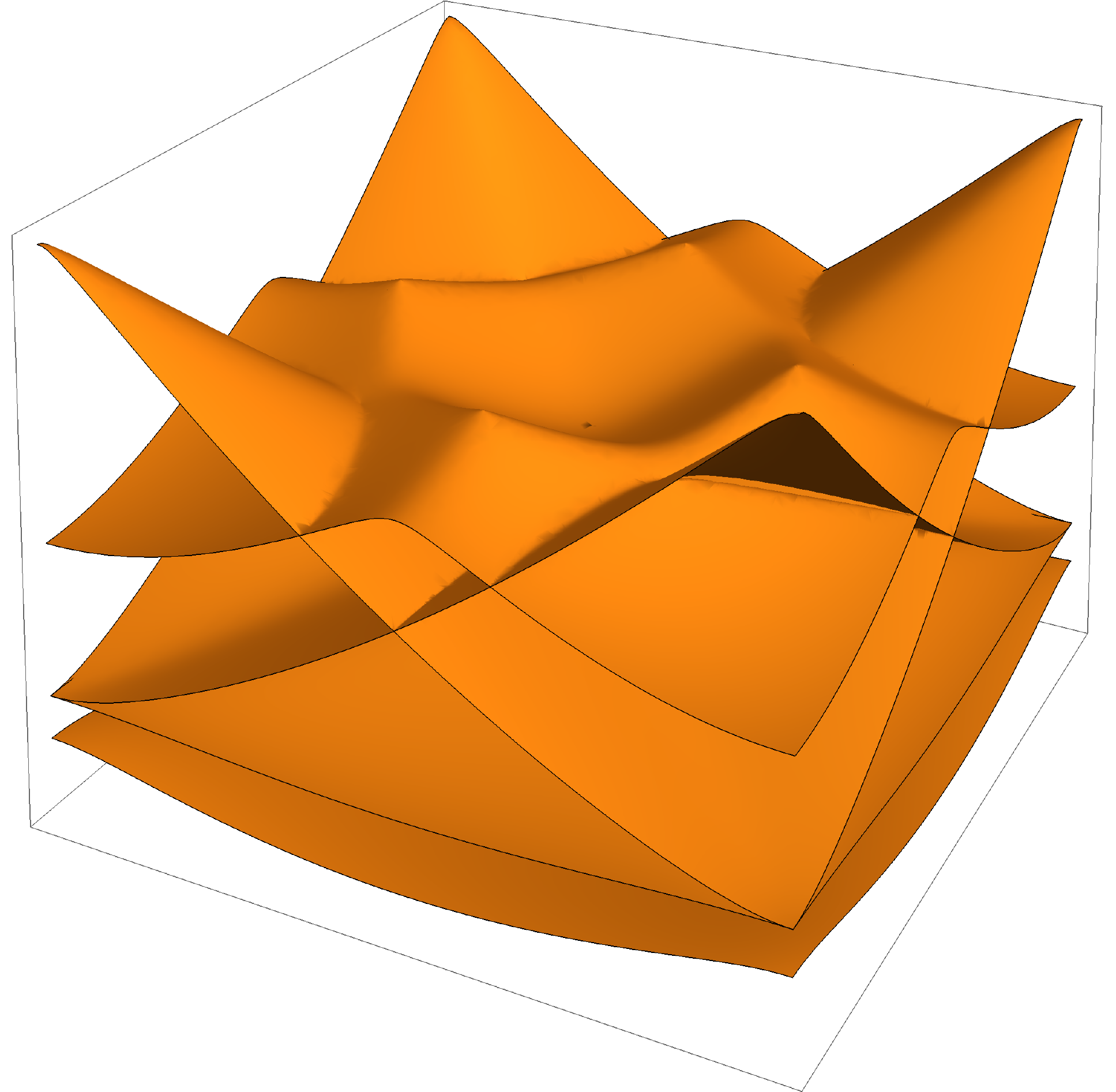}
  \caption{The electronic band structure or algebraic surface of a 2D toy model of a material. The band structure is often the integrand for integrals that give properties of materials. A band structure in 2D was chosen due to the difficulty of visualizing in more than 3D. The Brillouin zone in this case is simply a square.}
  \label{fig:bandstruct}
\end{figure}

Computing intrinsic properties of materials using density functional theory requires numerical integration \cite{cances2020numerical}. For example, the energy of the electrons (band energy) and the number of electrons in a given energy state (density of states) are properties of materials that are obtained by numerical integration. The domain of integration for these integrals is a Voronoi cell called the first Brillouin zone (referred to throughout this paper as simply Brillouin zone or BZ). A Voronoi cell is the region of space closer to one point in a mesh than to any other point in the mesh. In terms of geometry, the BZ is a convex polyhedron that often has a complicated shape. The integrand for the band energy or density of states is a periodic, algebraic surface called the electronic band structure. An example BZ for a 3D lattice is shown in Fig. \ref{fig:bz}, and the electronic band structure for a 2D toy model is shown in Fig. \ref{fig:bandstruct}.

The electronic band structure is computationally expensive to evaluate because each evaluation means solving an eigenvalue problem of a Hermitian matrix of order from hundreds to thousands. The number of evaluations is reduced by up to a factor of 48 by using the symmetry of the material, which allows one to reuse eigenvalues. In other words, if two points are symmetrically equivalent (for example, a rotation by $90^\circ$ maps one point to the other), the eigenvalues are identical at both points. 

Typically, the band energy is computed using the rectangular method despite its erratic and low-order error convergence \cite{jorgensen2021effectiveness}. The greater the symmetry of the integration grid, the greater the reduction in the number of eigenvalue computations. Calculating grids with the greatest possible symmetry has been an active area of research \cite{morgan2018efficiency}. The majority of the band energy computation is spent solving eigenvalue problems at points on a uniform grid over the BZ \cite{wende2019openmp}, so the savings from symmetry can be significant (as mentioned, up to 48 times more efficient).


\begin{figure}[t]
  \centering
  \includegraphics[width=\linewidth]{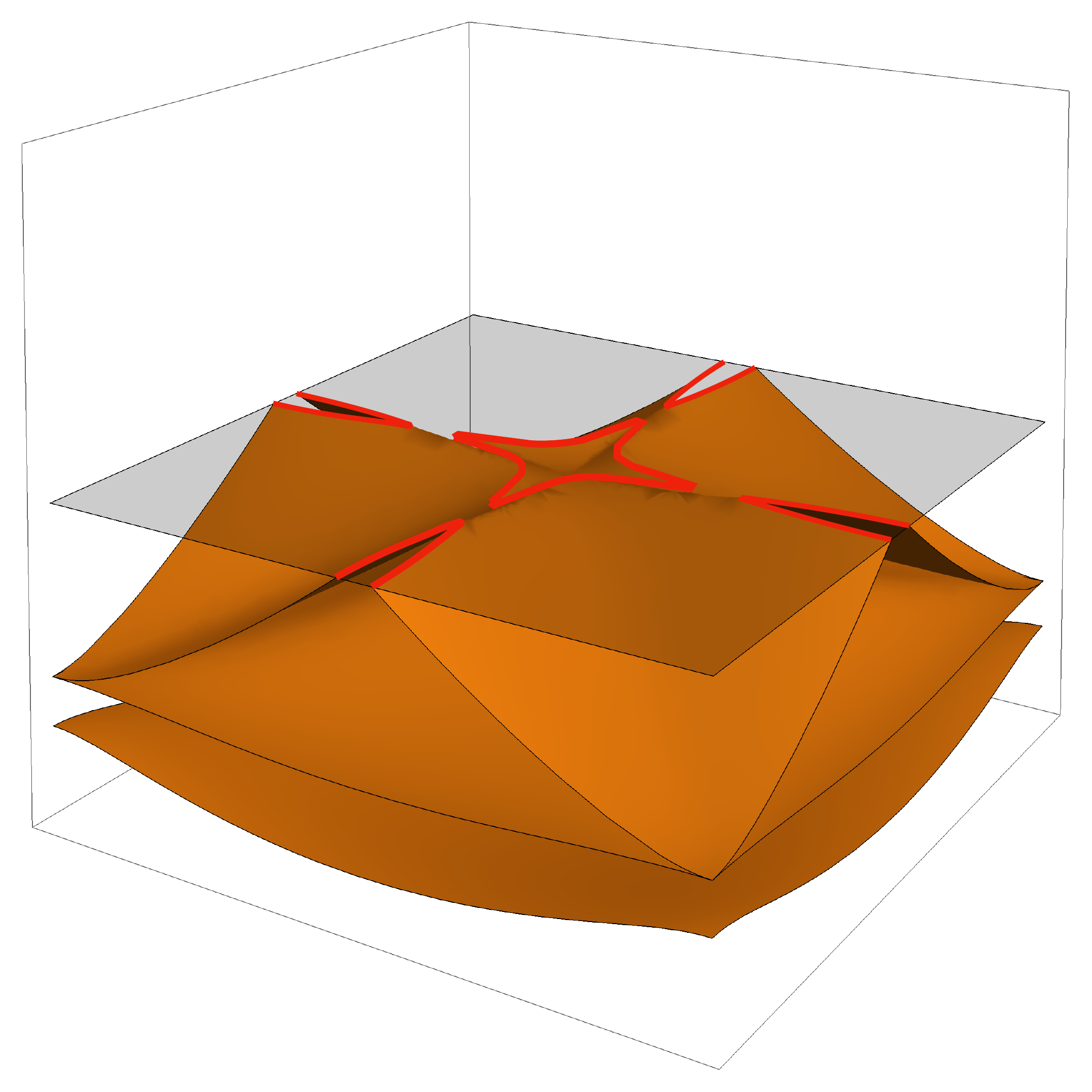}
  \caption{The band energy is the volume beneath the sheets of the algebraic surface below a cutting plane (shown in gray in the plot). The sheets are discontinuous where they intersect the cutting plane, and these discontinuities (highlighted in red) are the primary source of error in the calculation of the band energy.}
  \label{fig:pband-struct}
\end{figure}

However, uniform grids are not efficient because integration errors are not spread uniformly throughout the BZ but concentrated in small regions. This is demonstrated in Fig. \ref{fig:pband-struct} for the band energy calculation but is also the case for the calculation of other properties of materials. In the figure, most of the error is concentrated around the discontinuities ( integration errors are proportional to the height of the discontinuities), and a preferential sampling close to the discontinuities leads to improved integration error convergence.

Adaptive meshes are computationally more efficient but break symmetry (very few of the points in the mesh are equivalent to other points in the mesh by symmetry). Whereas symmetry may make uniform grids up to 48 times more efficient, symmetry does very little to improve the efficiency of adaptive meshes.

\begin{figure}[b]
  \centering
  \includegraphics[width=2in]{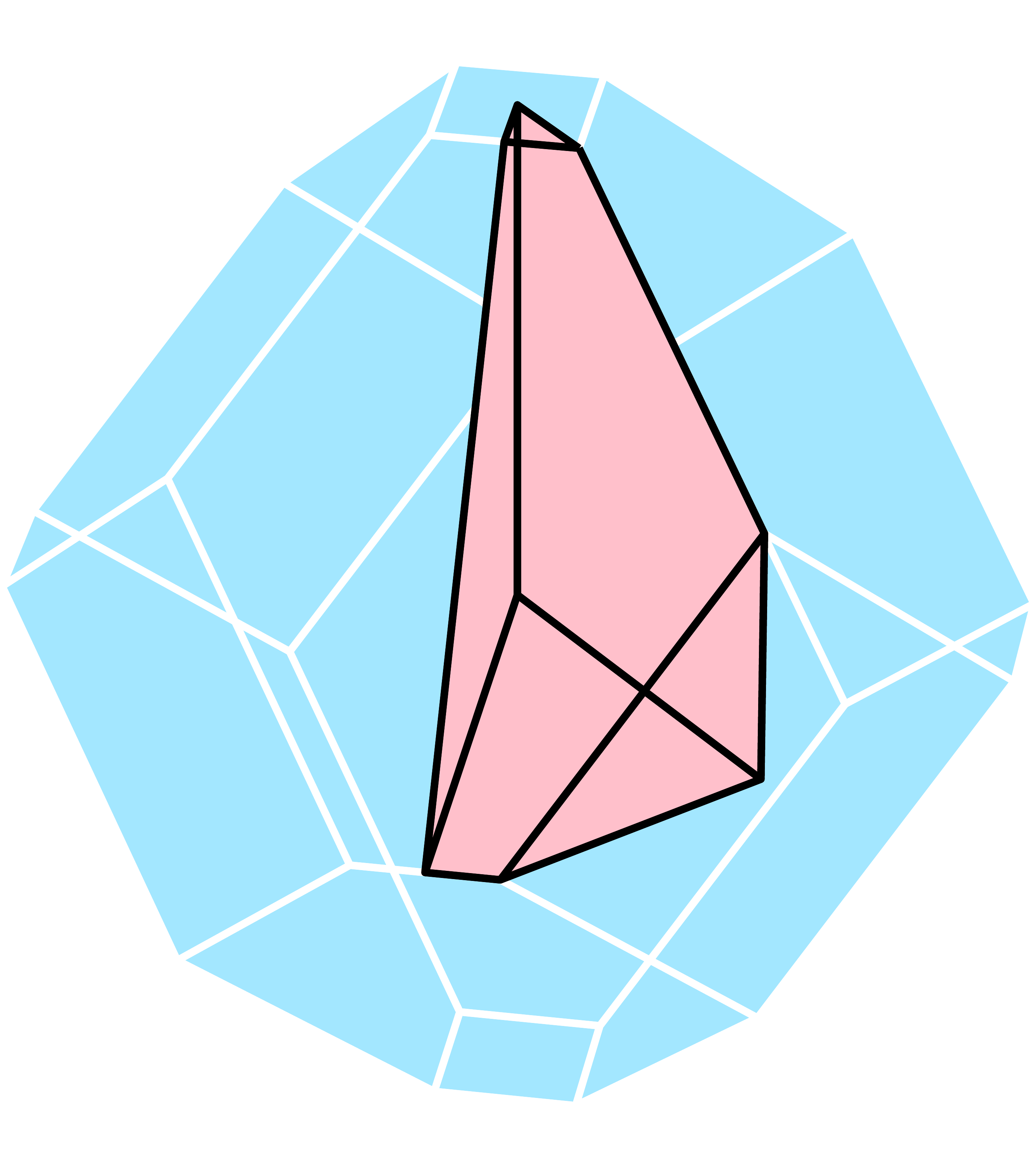}
  \caption{The Brillouin zone (in blue) and irreducible Brillouin zone (in pink) for a body-centered tetragonal lattice. The irreducible Brillouin zone is a convex polyhedron that often has a complicated shape.}
  \label{fig:comp-ibz}
\end{figure}

This apparent drawback of adaptive meshes is avoided by integrating solely within a part of the BZ called an {\it irreducible} Brillouin zone or IBZ. An IBZ is a closed polytope $Q$ within a BZ such that any lattice symmetry $g$ that moves any point of the interior of $Q$ must move $Q$ to a new polytope $gQ$, which only overlaps with $Q$ on the boundary, if at all.
By integrating within an IBZ, one has the freedom to refine regions where integration errors are large without taking an efficiency hit from breaking symmetry. Like the BZ, an IBZ is a convex polyhedron that often has a complicated shape (see Fig. \ref{fig:comp-ibz}).

Adaptive integration schemes split the domain of integration into subelements of various geometries. Because symmetry-breaking is no longer a concern when working in an IBZ, one has freedom to choose the shape of integration subelements (for example, simplices or hexahedra).

Another benefit of integrating within an IBZ is avoiding many of the intersections of the sheets of the algebraic surface, which typically occur at locations of high symmetry. Intersections of sheets in regions of high symmetry, called non-accidental crossings, occur at the boundary of an IBZ and have no effect on the accuracy of the interpolation of the band structure within an IBZ. \textit{Accidental} crossings, which occur \textit{within} an IBZ, are still problematic and affect the accuracy of interpolation significantly.

There are algorithms for calculating the BZ \cite{finney1979procedure}, a topic covered in nearly all solid-state physics textbooks. Despite many papers on calculating points and lines of high symmetry in the BZ \cite{munro2020improved}, we are aware of only one other algorithm \cite{otero2011fast} for calculating an IBZ, but it has little explanation and no results for verification.

In what follows, we present an algorithm that uses point symmetries to efficiently reduce the Brillouin zone to an irreducible Brillouin zone. The only inputs required are the lattice vectors, the atomic basis, and the crystal symmetries. A proof of the algorithm is provided in the appendix, and an implementation of the algorithm is available as open-source software. In summary, with a representation of an IBZ, properties of materials may be calculated with more efficient, higher-order, adaptive integration schemes because low-symmetry integration points do not affect the efficiency of the calculation.

\section{Calculating the Brillouin zone}
 
 \begin{figure}[t]
  \centering
  \includegraphics[width=\linewidth]{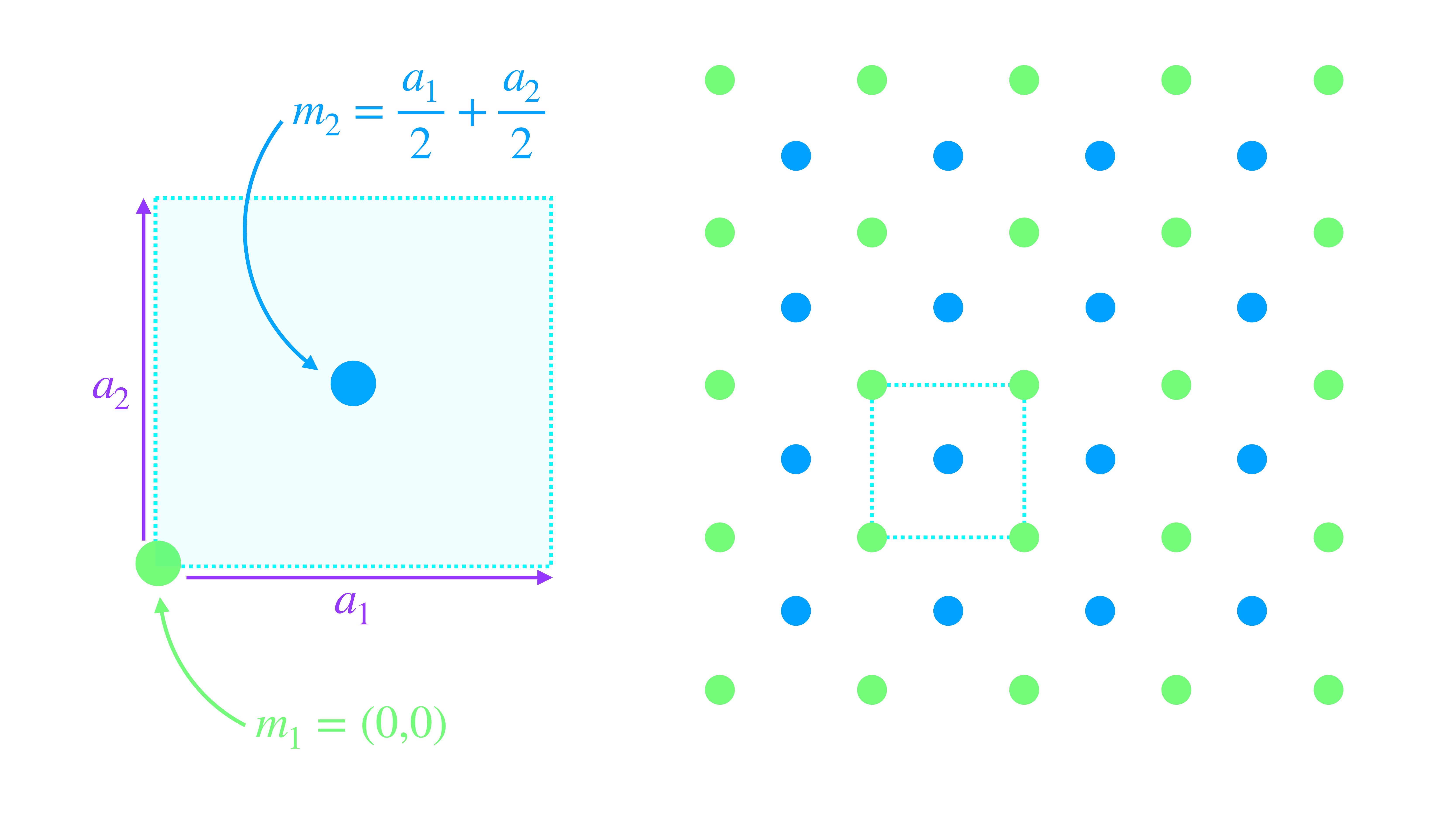}
  \caption{An example of a 2D crystal. Integer multiples of the lattice vectors $a_1$ and $a_2$ generate each point in the lattice. The positions of the atoms $m_1$ and $m_2$ with respect to the unit cell (the cyan square) are given in terms of the lattice vectors. The different colors of the dots indicate different atom types.}
  \label{fig:lat-unit}
\end{figure}
 
In many calculations, materials are treated as ideal crystals with atoms arranged periodically. The atoms of crystals lie on a lattice, an example of which is given in 2D in Fig. \ref{fig:lat-unit}.

\begin{figure}[t]
  \centering
  \includegraphics[width=\linewidth]{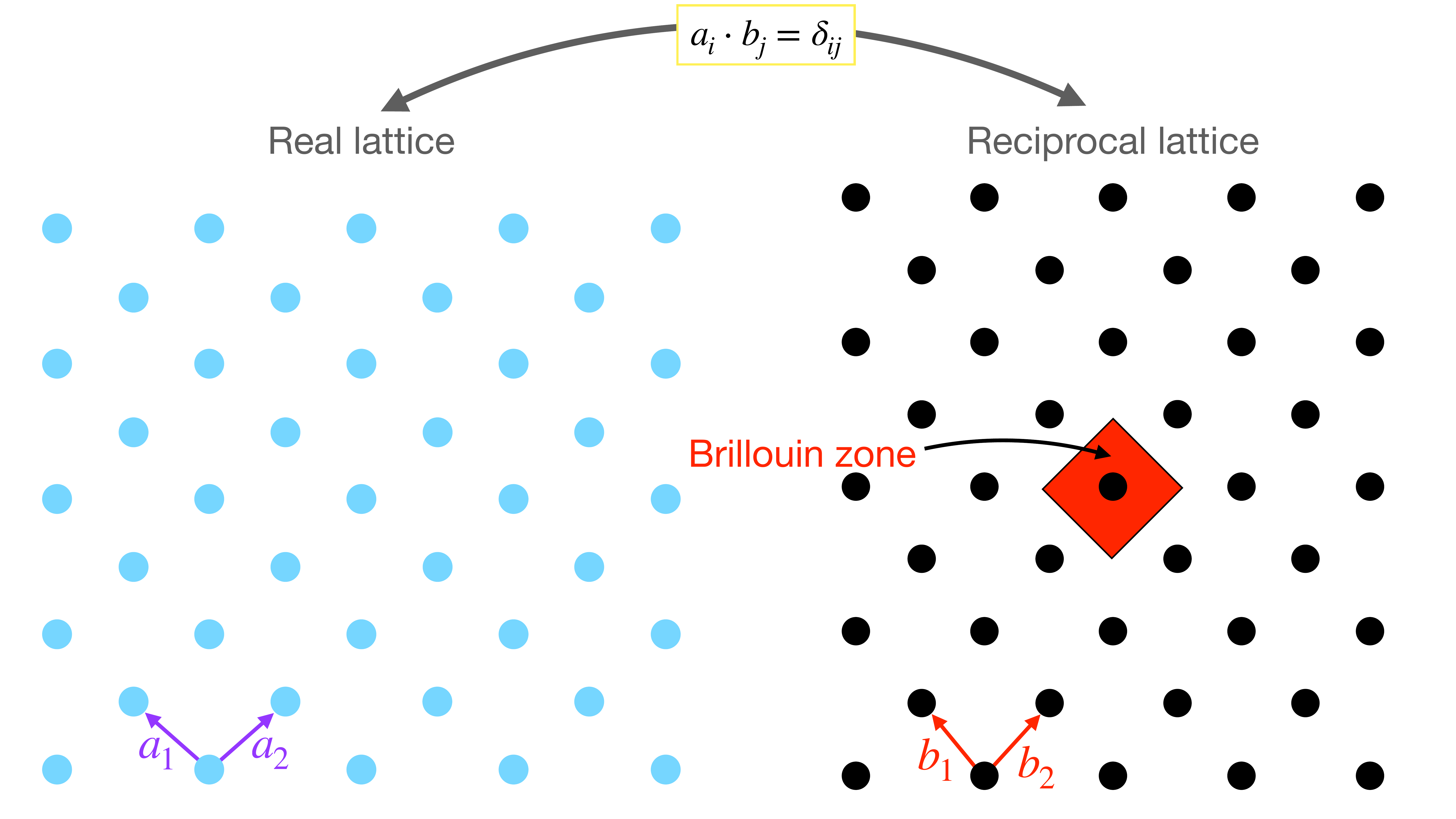}
  \caption{A Fourier transform maps the real-space lattice to the reciprocal lattice. The real and reciprocal lattice vectors satisfy the relation $a_i \cdot b_j = \delta_{ij}$ where $\delta_{ij}$ is the Kronecker delta. The Brillouin zone is one of infinite options for the unit cell of the reciprocal lattice and is the region of space closest to the origin of reciprocal space than any other point in the reciprocal lattice.}
  \label{fig:recip-lat}
\end{figure}

The arrangement of the atoms in an infinite crystal can be characterized by its lattice vectors and atomic basis 
\begin{equation}
x = A \mathbf{n} + \mathbf{m}_{\alpha}
\end{equation}
where $x$ is the position of an atom in the crystal, $A$ is a matrix whose columns are the lattice vectors (in 2D, $a_1$ and $a_2$), $\mathbf{n}$ is a vector of integers, and $\mathbf{m}_{\alpha}$ is the offset of the $\alpha$th atom in the unit cell. The reciprocal lattice, obtained by taking the Fourier transform of the real-space lattice, satisfies the relation $A B = I$ (there may be a factor of $2\pi$, depending on the convention) where $I$ is the identity matrix and $B$ is a matrix whose columns are the reciprocal lattice vectors ($b_1$ and $b_2$ in Fig. \ref{fig:recip-lat}). A lattice may be thought of as a tessellation of $\mathbb{R}^n$ where the tile of the tessellation is called the unit cell. For a single lattice, there are an infinite number of possible tiles that can tesselate $\mathbb{R}^n$. The BZ is a unit cell comprising the region of reciprocal space closest to the origin than to any other reciprocal lattice point. These concepts are illustrated in Fig. \ref{fig:recip-lat}.

\begin{figure}
  \centering
  \includegraphics[width=\linewidth]{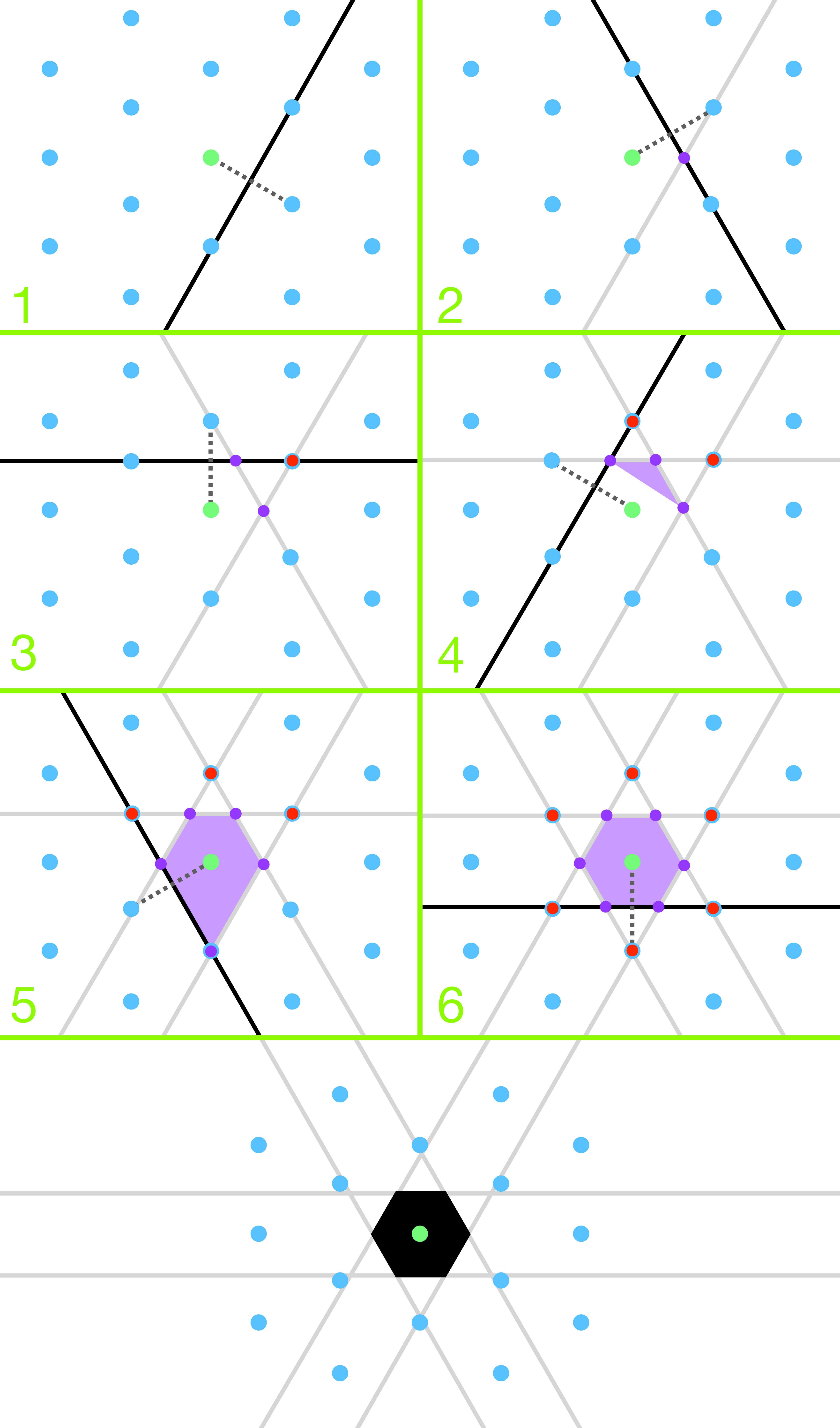}
  \caption{An illustration of the typical algorithm for computing the Brillouin zone. The reciprocal lattice points are blue dots, the origin is the green circle, line segments connecting the origin to lattice points are dashed lines, the perpendicular bisectors are the solid black or gray lines, intersections of bisectors that are kept are violet dots, and intersections that are discarded are red dots. After each iteration, a candidate BZ is constructed (if possible) from intersections of bisectors, shown as violet polygons in the figure.}
  \label{fig:bz-compare}
\end{figure}

Constructing the BZ is a topic almost always treated in introductory solid state physics textbooks. The algorithm is typically described as follows:
\begin{itemize}
  \setlength\itemsep{0em}
  \item Calculate a few lattice points \footnote{This is explained in greater detail later on.} near the origin.
  \item Order the lattice points by distance from the origin.
  \item For each lattice point near the origin:
  \begin{itemize}
    \item[-] Determine the straight line segment that connects the origin to the lattice point.
    \item[-] Calculate the perpendicular bisector of the line segment (this is a straight line that intersects the line segment at its midpoint and is perpendicular to the line segment).
    \item[-] Calculate intersections of the perpendicular bisector with any previously calculated perpendicular bisectors.
    \item[-] Discard any intersection that is on the side of any perpendicular bisector that is opposite the origin.
    \item[-] Break out of the loop if a convex hull made from the intersections has the same size as $\det(B)$.
  \end{itemize}
\end{itemize}
In this approach, one has to compute intersections of perpendicular bisectors and discard intersections that are opposite the origin of any bisector. This becomes more complicated in 3D where the bisectors are planes instead of lines, and intersections of bisecting planes are lines instead of points. The algorithm is illustrated in Fig. \ref{fig:bz-compare}.

\begin{figure}
  \centering
  \includegraphics[width=\linewidth]{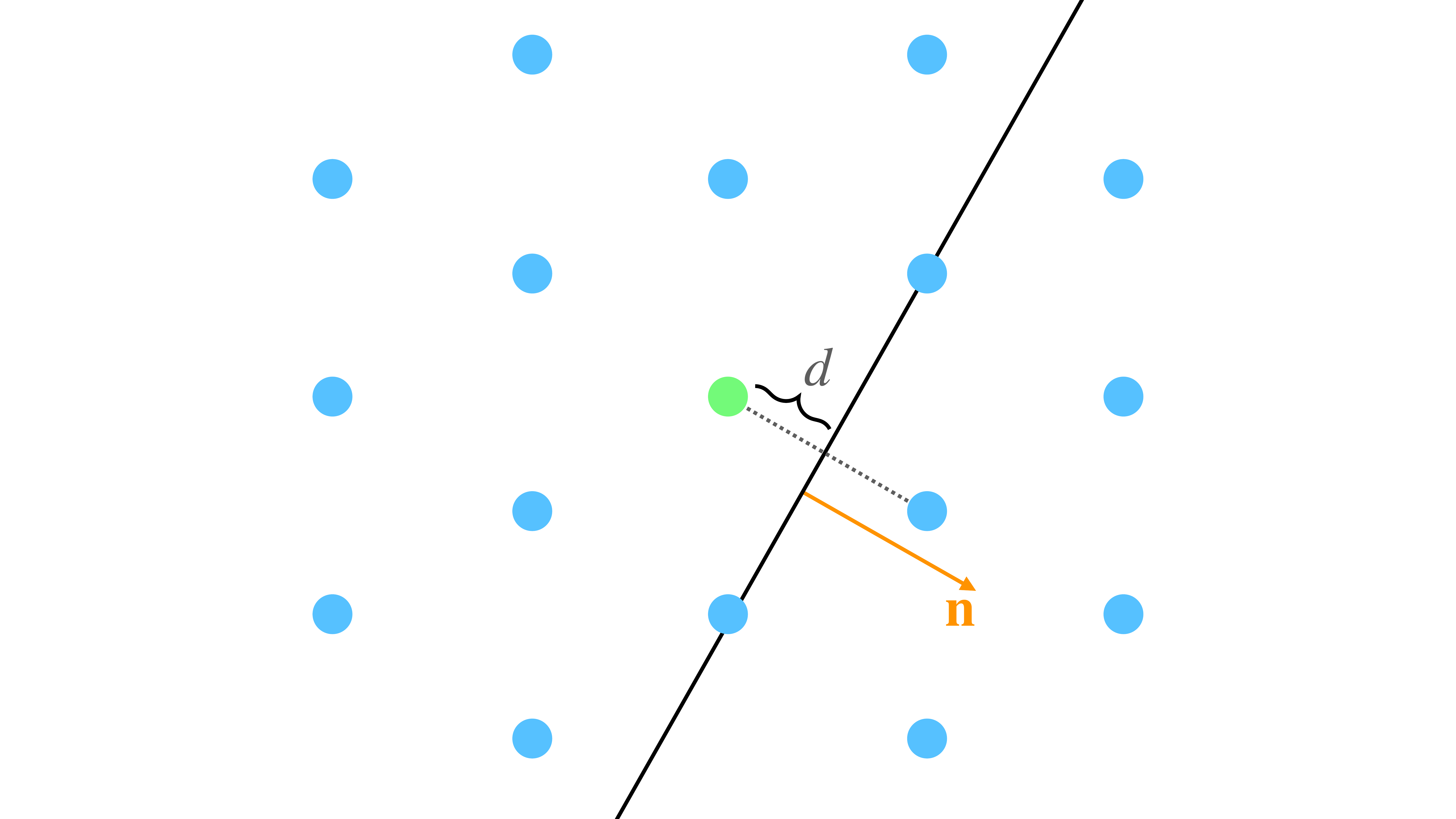}
  \caption{A half-space is a tuple $(\vec{n},d)$ where $\vec{n}$ is a unit vector normal to the perpendicular bisector (the orange arrow in the figure) and $d$ is the distance from the origin (green circle) to the bisector (solid black line) in the direction of $\vec{n}$.}
  \label{fig:half-space-ill}
\end{figure}

The algorithm is simplified significantly by taking advantage of half-spaces. A half-space is a tuple $(\vec{n},d)$ where $\vec{n}$ is a unit vector that is normal to the perpendicular bisector (in this case), and $d$ is the distance from the origin to the bisector in the direction of $\vec{n}$. This is illustrated in Fig. \ref{fig:half-space-ill}.

\begin{figure}[h]
  \centering
  \includegraphics[width=\linewidth]{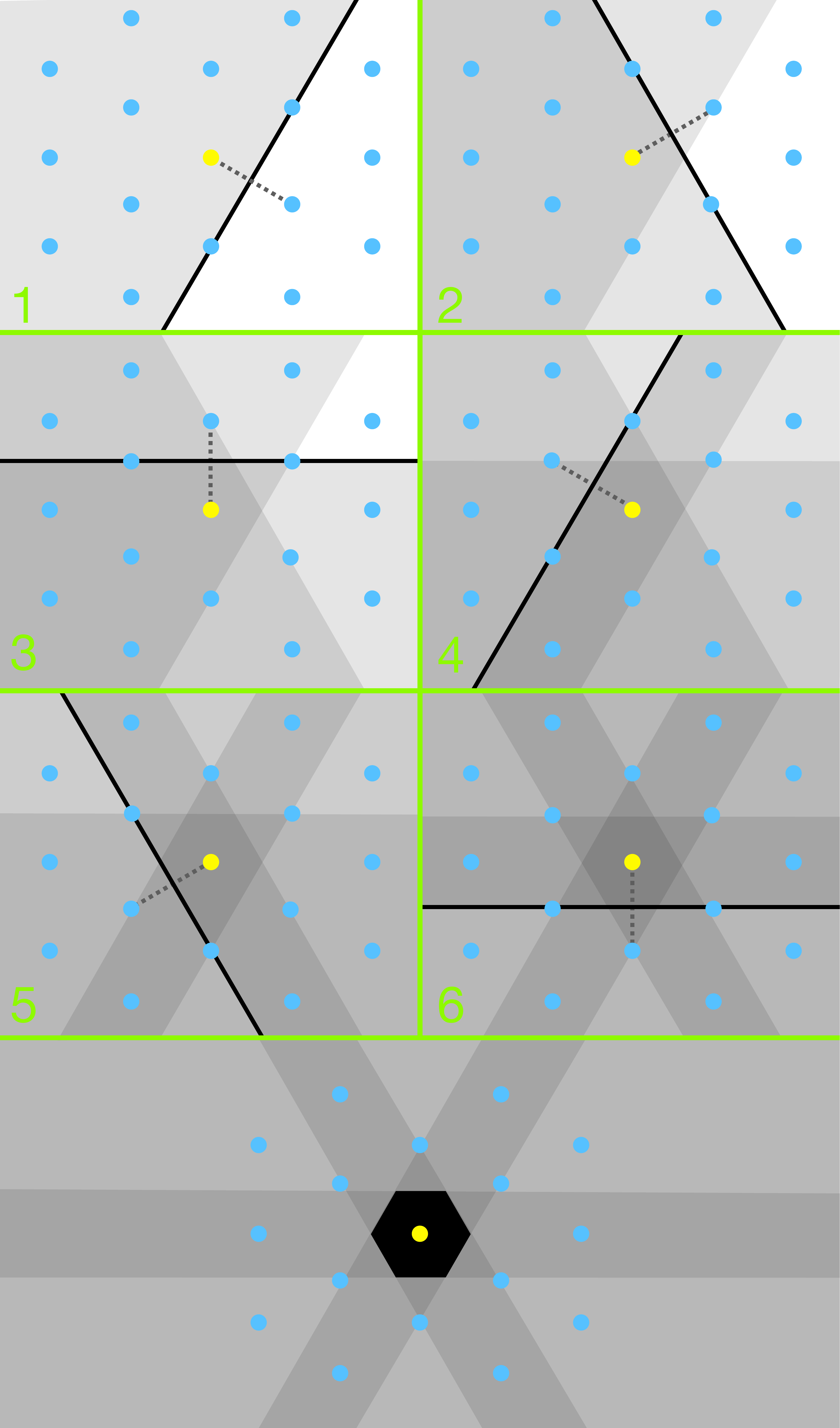}
  \caption{A simpler algorithm for computing the Brillouin zone involving half-spaces. The origin is the yellow circle, the reciprocal lattice points near the origin are blue dots, and half-spaces are the gray-shaded regions. The intersection of half-spaces gives the Brillouin zone.}
  \label{fig:bz-compare-int}
\end{figure}

With half-spaces, one does not need to calculate intersections of bisectors nor keep track of intersections. The algorithm for calculating the BZ becomes:

\begin{itemize}
  \setlength\itemsep{0em}
  \item Calculate a few lattice points near the origin.
  \item For each lattice point near the origin:
  \begin{itemize}
    \item[-] Calculate the half-space for the point (the distance is the norm of the point divided by 2, and the normal vector is in the same direction as the vector that points from the origin to the lattice point).
    \item[-] Calculate the intersection of the half-space with any previously calculated half-spaces.
    \item[-] Break out of the loop if the size of the intersection of all the half-spaces is the same as $\det(B)$.
  \end{itemize}
\end{itemize}
This approach to calculating the BZ is illustrated in Fig. \ref{fig:bz-compare-int}. A more detailed outline of the algorithm that gives the BZ is provided below.

\begin{algorithm}[H]
  \caption{Construct the BZ}\label{alg:BZ}
  \begin{algorithmic}[1]
  \Procedure{Construct\_BZ}{$\B$}\\
  \Comment{$\B$ is a matrix with reciprocal lattice vectors as columns}
  \State $\V \gets$ Reciprocal lattice points near the origin.
  \State $\M \gets$ Initialize the set of half-spaces.
  \For{$v \in \V$}
  \State{$H_{v} \gets \{x\in \R^n | x \cdot \frac{v}{|v|} \leq \frac{|v|}{2}\}$}\\
  \Comment{The half-space bisecting the line segment from $O$ to $v$}
  \State{$\M \gets \M \cup \{H_{v}\}$} \Comment{Insert $H_{v}$ into $\M$.}
  \EndFor
  \State \Return{$\M$}
  \EndProcedure
  \end{algorithmic}
\end{algorithm}

\begin{figure}[t]
  \centering
  \includegraphics[width=\linewidth]{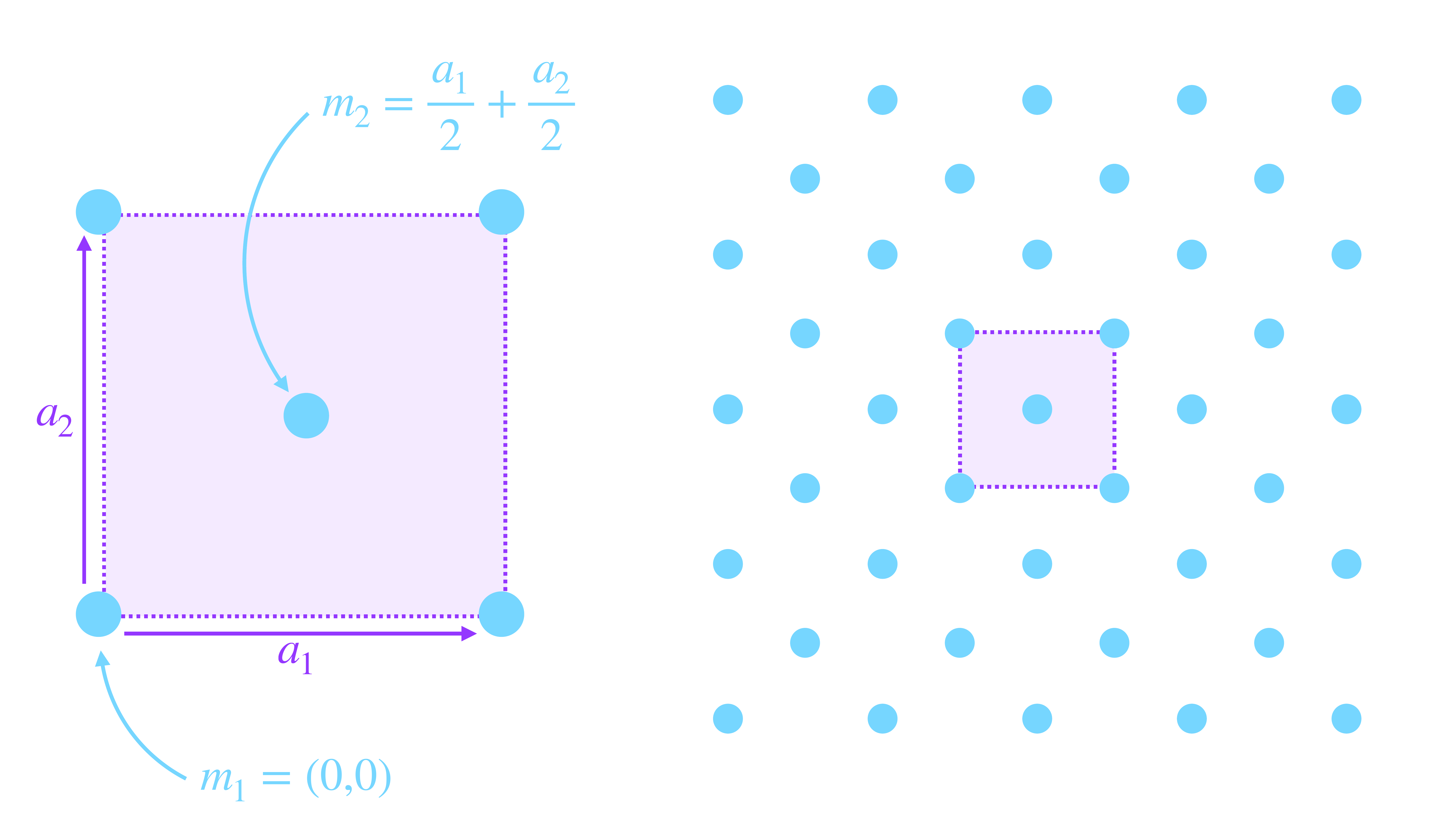}
  \caption{A non-primitive unit cell for a crystal with square symmetry.}
  \label{fig:lat-unit-prim} 
\end{figure}

\begin{figure}[h]
  \centering
  \includegraphics[width=\linewidth]{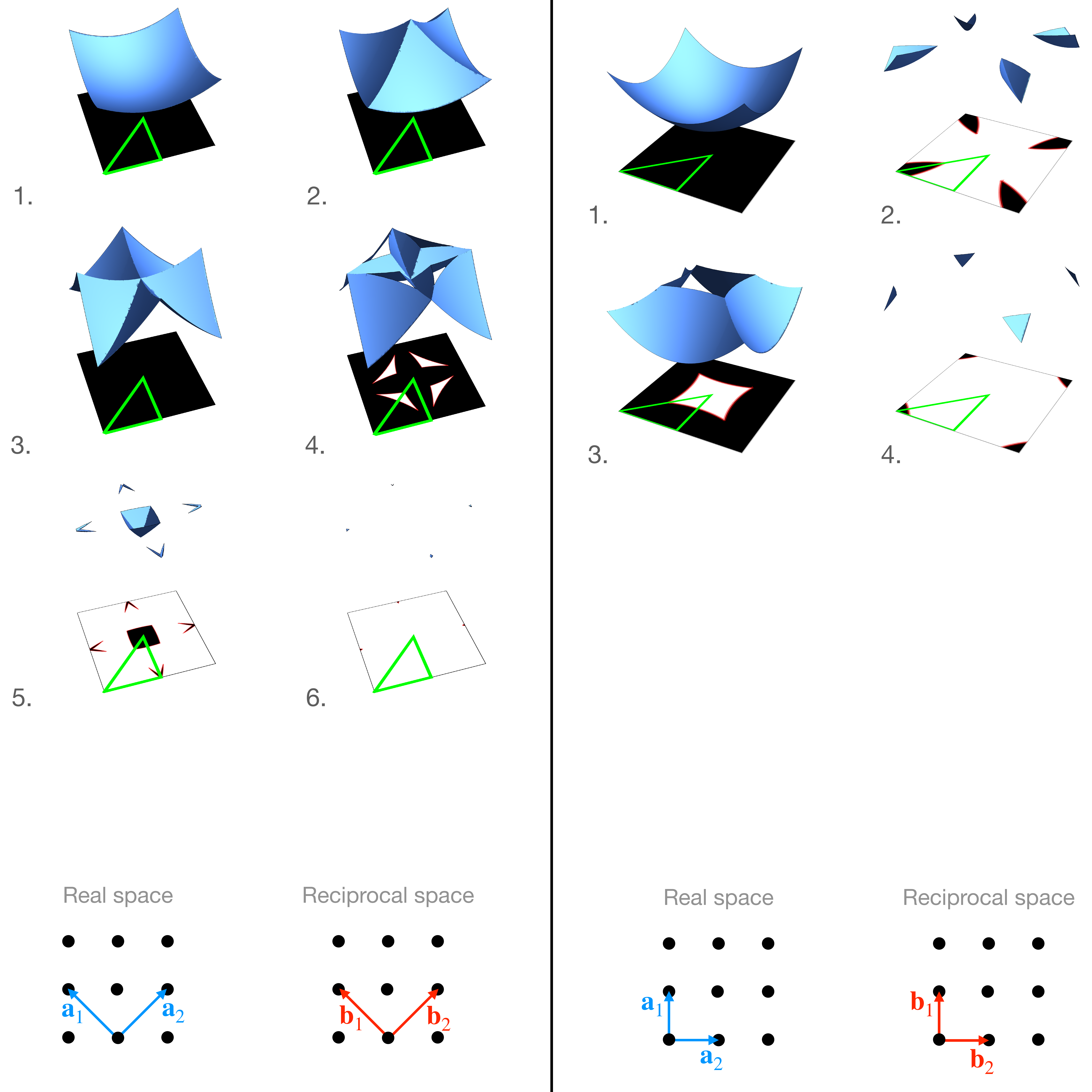}
  \caption{The sheets of the algebraic surface on the left side of the figure are for a non-primitive basis (lattice vectors) for a 2D free-electron model (often used to model metallic materials), while the sheets on the right are for a primitive basis. Below each sheet is the shadow of the sheet or the domain of integration for the band energy calculation as well as the IBZ outlined in green. Calculating the band energy with a non-primitive is much more difficult because the integration is over more sheets, and the sheets of the algebraic surface have more self-intersections that create difficult-to-integrate, fine features. Sharp corners in the shadows of the sheets are very difficult to approximate when they are interior to the IBZ, which are present for the non-primitive basis on the left.}
  \label{fig:prim-compare}
\end{figure}

The unit cell for a crystal that is as small as possible and contains only one lattice point is called a \textit{primitive} unit cell. Fig. \ref{fig:recip-lat} is an example of a primitive unit cell. The same example with a non-primitive unit cell is shown in Fig. \ref{fig:lat-unit-prim}. It is beneficial to make the unit cell primitive because the integral in the calculation of properties of materials is over fewer sheets of the algebraic surface, and the algebraic surface has fewer self-intersections. Several codes are available for making unit cells primitive \cite{ong2013python,morgan2021symlib}. In the case of the band energy calculation, self-intersections introduce very fine features that are difficult to capture without large numbers of integration points. This is shown in Fig. \ref{fig:prim-compare}

\begin{figure}[h]
  \begin{center}
  \includegraphics[width=0.7\linewidth]{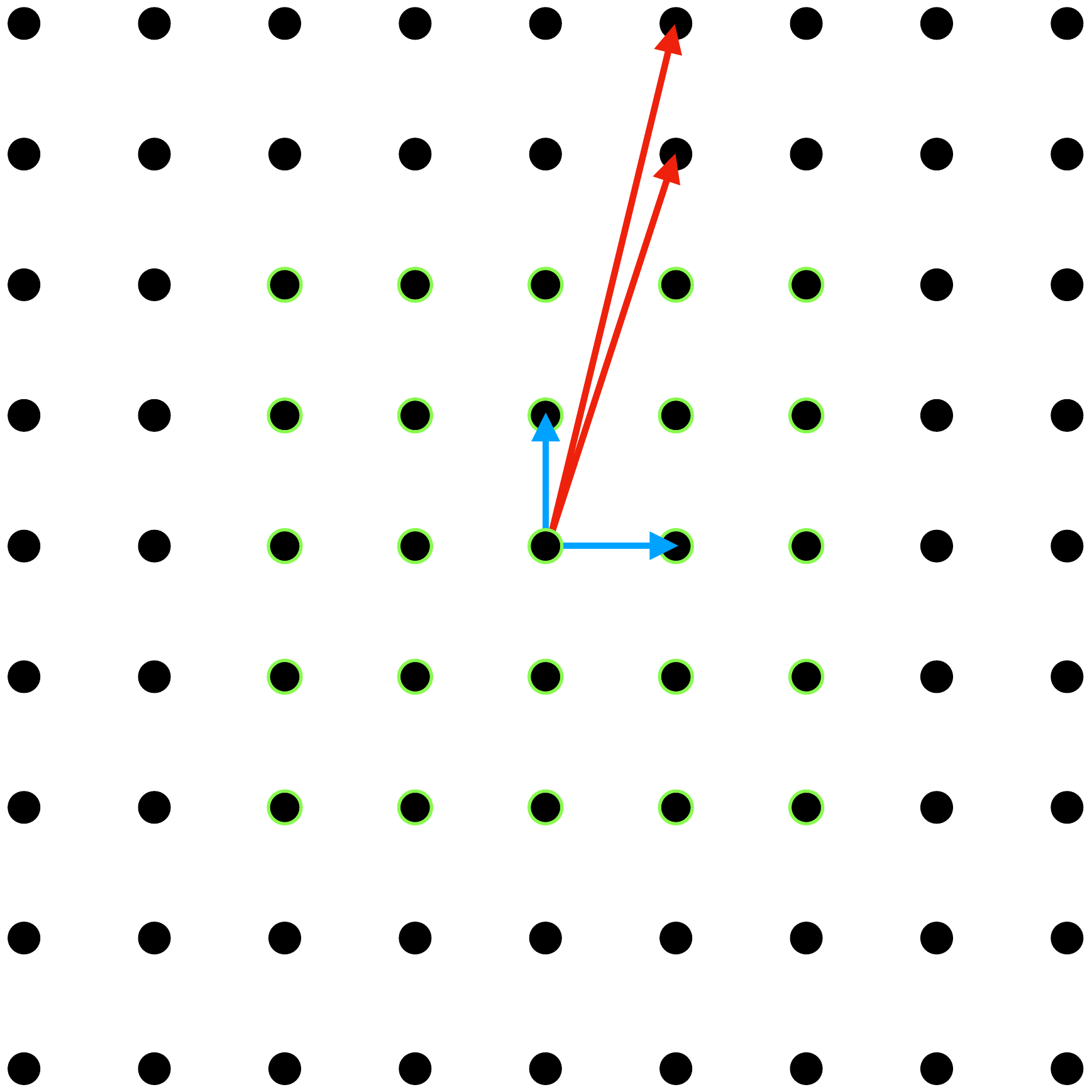}
  \caption{The lattice vectors in red and blue both produce the same lattice, shown in black. In general. the lattice points outlined in green are the only lattice points needed to compute the BZ and often only a subset of these points are necessary. For the lattice vectors that have not been Minkowski reduced (shown in red), far more lattice points (large integer values for $i$ and $j$ in Eq. \ref{eq:latpts-2d}) have to be considered to generate the lattice points outlined in green.}
  \label{fig:mink-reduce}
  \end{center}
  \end{figure}

In Algorithm \ref{alg:BZ}, it was left unspecified how many reciprocal lattice points are needed; one might not include enough to determine the correct BZ or include far too many. (Usually, the number of lattice points is small, and this is not an issue except when the reciprocal unit cell is very skew. See Fig. \ref{fig:mink-reduce} for an example in 2D.) In 2D and 3D, as long as the reciprocal lattice vectors are Minkowski reduced \footnote{Minkowski reduction makes the lattice vectors as short as possible. See \protect \onlinecite{nguyen2009low} for details.} the BZ is guaranteed to lie within the set of unit cells that share a vertex at the origin \footnote{See the appendix of \protect \onlinecite{hart2019robust}.} This puts constraints on the number of reciprocal lattice points that have to be included in the calculation of the BZ. To be specific, the lattice points that are sufficient to determine the BZ in 2D are
\begin{equation} \label{eq:latpts-2d}
x = B \begin{pmatrix} i \\ j \end{pmatrix},  \qquad i,j \in \{-2,-1,0,1,2\},
\end{equation}
where $B$ is a matrix with the reciprocal lattice vectors as columns. In 3D, there is one more iterator over the same range, that is, 
\begin{equation}
x = B \begin{pmatrix} i \\ j\\ k \end{pmatrix},  \qquad i,j,k \in \{-2,-1,0,1,2\}.
\end{equation}
Minkowski reduction makes the algorithm robust because it specifies the number of reciprocal lattice points needed.

  \begin{figure}[h]
  \begin{center}
  \includegraphics[width=0.8\linewidth]{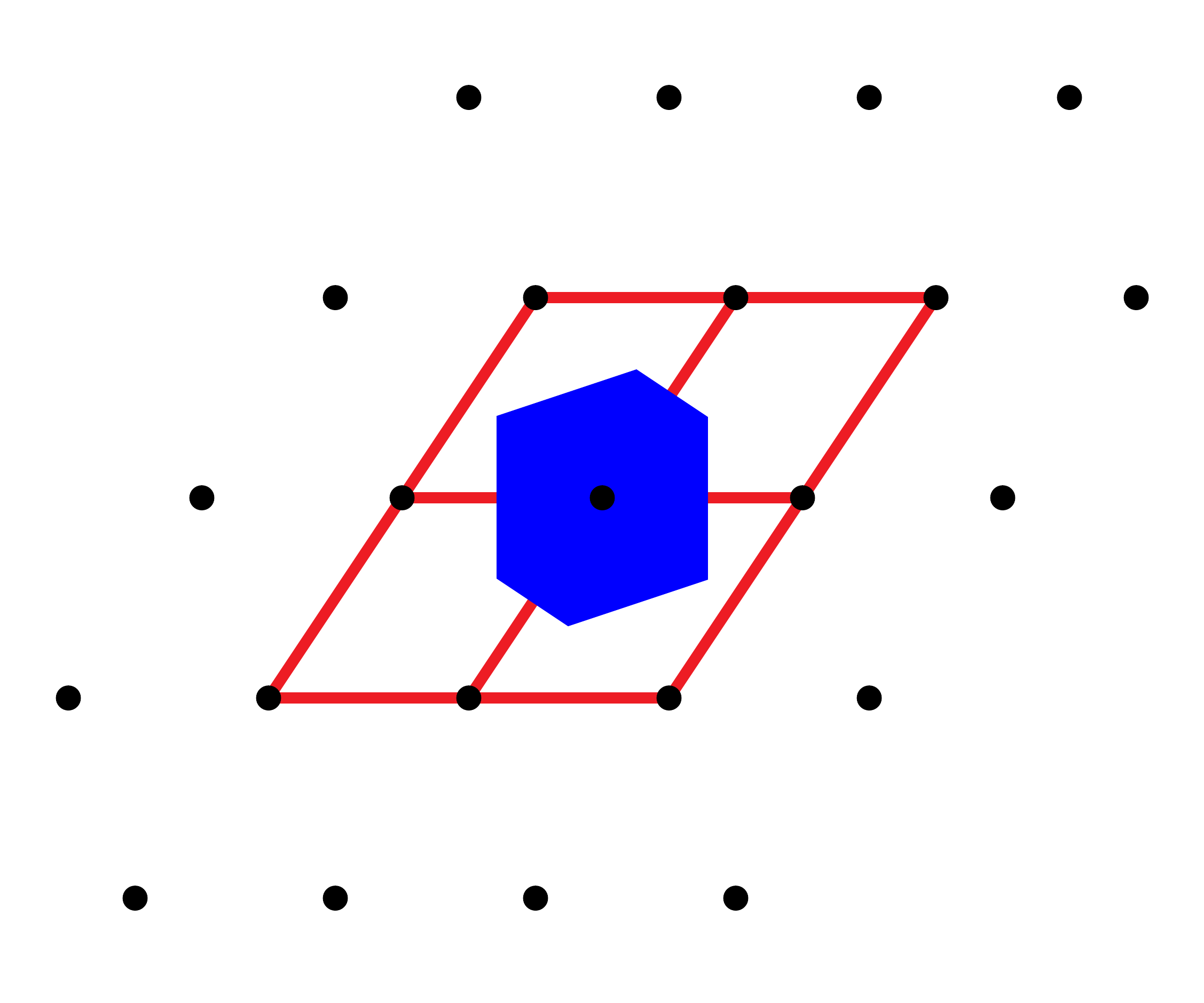}
  \caption{The Brillouin zone of a 2D lattice lies within the conventional unit cells that have a vertex at the origin when the lattice basis has been Minkowski reduced \protect \footnote{A basis is Minkowski reduced when the lattice vectors are as short as possible. See \protect \onlinecite{nguyen2009low}.}. Reciprocal lattice points are shown as black points, the unit cells are outlined in red, and the Brillouin zone is shaded in blue.}
  \label{fig:bz-boundary}
  \end{center}
  \end{figure}

\section{Calculating an irreducible Brillouin zone}

The following is the algorithm for generating an irreducible Brillouin zone. Although the basic concepts of the construction for the BZ are well known and discussed in most textbooks, they do not provide a practical representation for the BZ or a complete description of the geometry of the BZ (vertices, edges, faces, and volume). To the best of our knowledge, prior to our algorithm, a rigorous and general algorithm for calculating a  point-symmetry reduced Brillouin zone (an irreducible Brillouin zone) and characterizing its geometric features has not been developed.

The point symmetries of the real-space crystal are used to reduce the BZ to an IBZ as follows. We define $\G$ to be the point group of the crystal's space group (the space group includes rotational and translational symmetries), $\I \in \G$ is the identity operator, $\P$ is the BZ convex polyhedron, and $\V$ is the set of vertices of $\P$. An outline of the algorithm that reduces the BZ to an IBZ is given below (for $x,y\in \R^n$ the symbol $d(x,y)$ denotes the distance between $x$ and $y$). See the Appendix for more details.
  
\begin{figure}[h]
  \begin{center}
  \includegraphics[width=\linewidth]{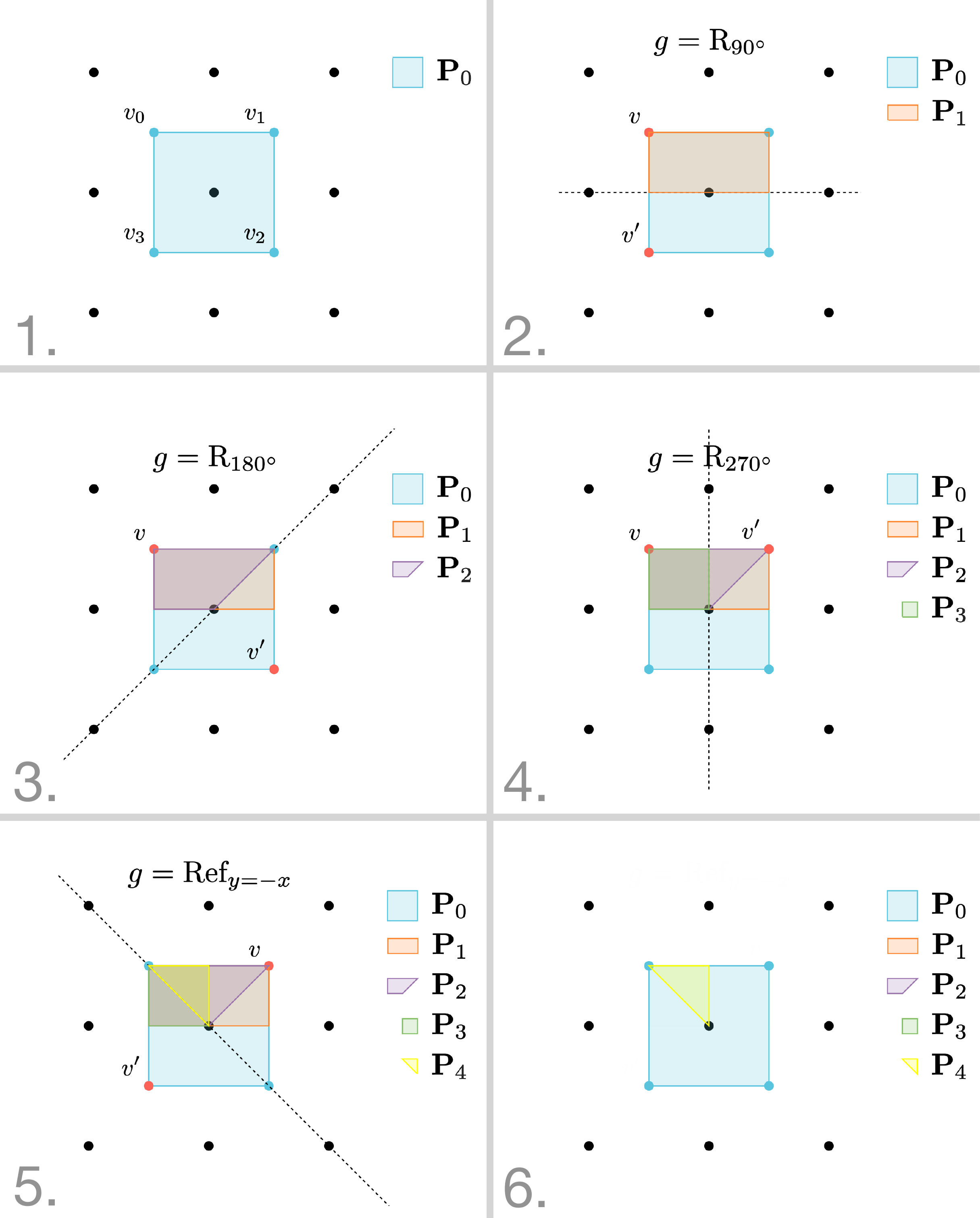}
  \caption{An illustration of the IBZ calculation. We begin with the BZ labeled $P_0$. Each operator in the point group successively reduces the size the BZ. Moving down the figure, we show the reduction of the BZ resulting from selected operators from the point group and end with the IBZ labeled $P_4$.}
  \label{fig:algorithm}
  \end{center}
\end{figure}

Reducing the BZ to an IBZ is shown in Fig.~\ref{fig:algorithm}. When the algorithm finishes, the resulting IBZ is the intersection of the half-spaces in $\N$ (see Algorithm 2), which is returned as a convex hull object. A convex hull object is a convenient way to store vertices, edges, and faces of the IBZ. 

We have written the IBZ algorithm in Julia \cite{bezanson2017julia} with dependencies on the Julia Polyhedra library \cite{poly} and the C qhull package \cite{qhull}. The code includes functions for visualizing the BZ and IBZ (many of the figures in this paper were produced from it) and can be called from Python. 

\begin{algorithm}[H]
  \caption{Construct an IBZ from $\P$ (the BZ)}\label{alg:IBZ}
  \begin{algorithmic}[1]
  \Procedure{Construct\_IBZ}{$\M$,$\G$}\\
  \Comment{$\M$ is a set of half-spaces whose intersection is $\P$.}\\
  \Comment{$\G$ is the (finite) symmetry group of $\P$. \phantom{rsecinBZ.}}
  \State $\V \gets$ The set of vertices of $\P$.
  \State $\F \gets \G \setminus \{\I\}$ \Comment{Initialize the set of symmetries.} 
  \State $\N \gets \makecopy{\M}$ \Comment{Initialize the set of half-spaces.} 
      \For{$v \in \V$}
        \For{$g \in \F$}
            \State $v' \gets g v$ \Comment{Calculate the transformed point.}
            \If{$v' \neq v$}
              \State{$H_{v,g} \gets \{x\in \R^n | d(x,v) \le d(x,gv)\}$}\\
                 \Comment{The half-space bisecting the line segment from $v$ to $v'$}
              \State{$\N \gets \N \cup \{H_{v,g}\}$} \Comment{Insert $H_{v,g}$ into $\N$.}
              \State{$\F\gets \F \setminus \{g\}$} \Comment{Remove $g$ from $\F$.}
            \EndIf
        \EndFor
      \EndFor
  \State \Return{$\N$}
  \EndProcedure
  \end{algorithmic}
  \end{algorithm}

\section{Testing the Implementation}

\begin{figure}[t]
  \begin{center}
  \includegraphics[width=\linewidth]{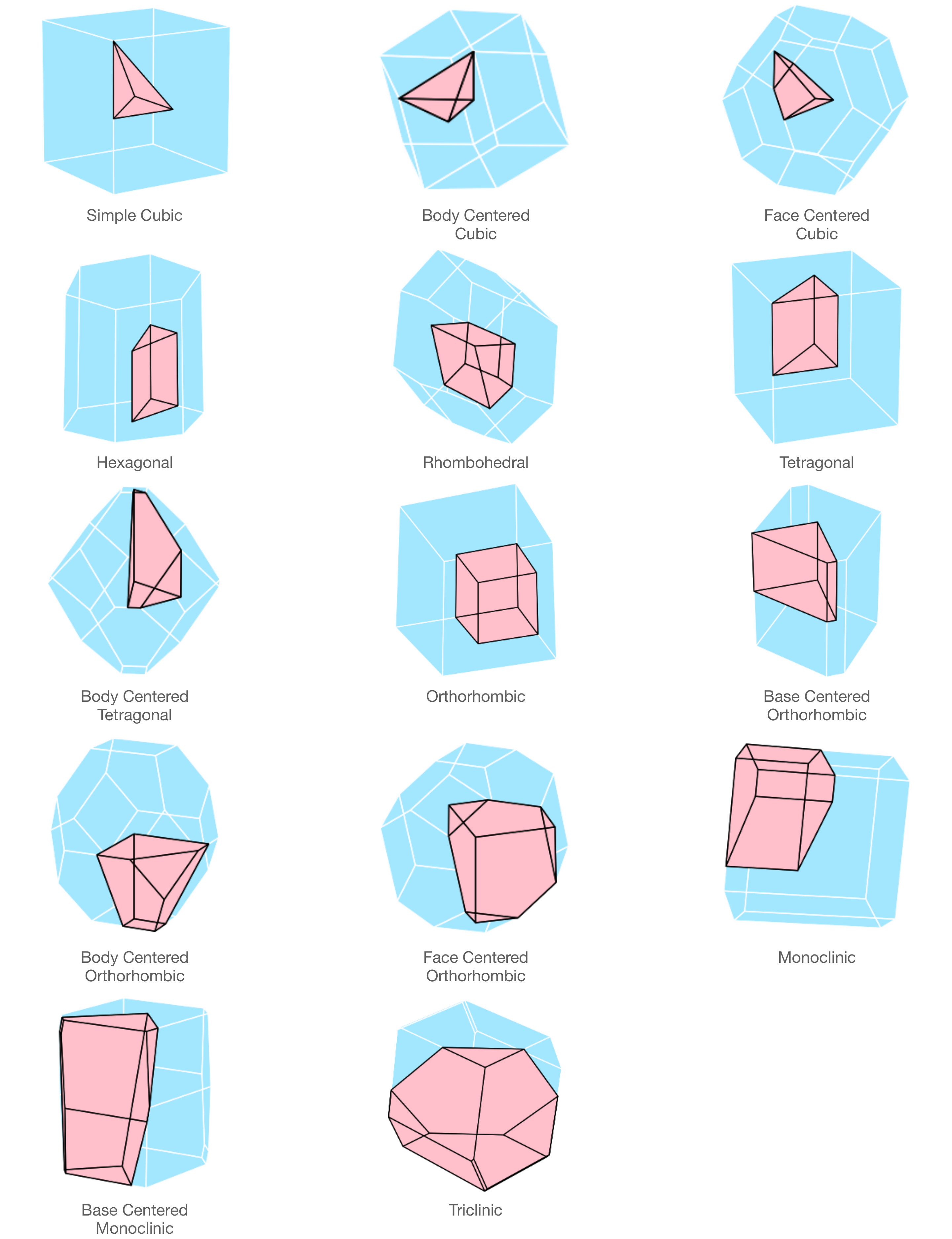}
  \caption{One BZ and IBZ for each of the 3D Bravais lattices.}
  \label{fig:combined-ibz}
  \end{center}
  \end{figure}

We know beforehand the relationship\cite{aroyo2014brillouin} between the IBZ volume $\operatorname{Vol}_{\mathrm{IBZ}}$, the BZ volume $\operatorname{Vol}_{\mathrm{BZ}}$, and the size of the point group $n_p$:
\begin{equation}
\operatorname{Vol}_{\mathrm{IBZ}} = \frac{\operatorname{Vol}_{\mathrm{BZ}}}{n_p}.
\end{equation}
This relation is verified for each IBZ calculation. For testing, we \textit{unfold the IBZ}. The IBZ is unfolded by applying each operator $g \in G$ to each of the vertices of the IBZ. If the algorithm is working correctly, the unfolded vertices are the same as the vertices of the BZ. Together these two calculations guarantee the IBZ is correct. The second step is necessary because it is possible to get the correct volume reduction but have the wrong shape. The IBZ is not unique, and the IBZ obtained from the algorithm depends on the order of the BZ vertices $V$ and the order of the point operators $G$. In some cases, faces (3D) or edges (2D) of the IBZ may be translationally or rotationally equivalent (see the appendix for a discussion). Plots of the results of the IBZ algorithm for each of the 14 Bravais lattices are shown in Fig. \ref{fig:combined-ibz} (We only show one BZ and IBZ even though some Bravais lattices have multiple BZs \cite{setyawan2010high}; we would also obtain many more IBZs for atomic bases that break the symmetry of the lattice).

\onecolumngrid
\onecolumngrid

  \twocolumngrid
  

\FloatBarrier
\section{Summary}

We have developed an algorithm that can compute an irreducible Brillouin zone (IBZ) of a crystal using the lattice vectors, atomic basis, and crystal symmetries. Working within an IBZ for calculations of properties of materials provides more freedom in selecting points for BZ integrations than uniform grids over the BZ because the grids no longer have to preserve symmetry. In particular, it allows the use of adaptive grids with high sample point densities at locations of high integration error. The calculation of an IBZ is simplified by using convex hull and half-space representations of the IBZ, which makes many aspects of the BZ reduction trivial.

\section{Declarations}

\subsection{Funding}
This work was supported by ONR (MURI N00014-13-1-0635).

\subsection{Conflicts of interest/Competing interests}
The authors have no conflicts of interest that are relevant to the content of this article.

\subsection{Availability of data and material}
Data sharing not applicable to this article as no datasets were generated or analyzed during the current study.

\subsection{Code availability}
The algorithm is a registered Julia package \href{https://github.com/jerjorg/SymmetryReduceBZ.jl}{\textsc{SymmetryReduceBZ}} that can also be called from Python.

\section{Appendix}
    \label{sec:proof}

Here we prove that Algorithm~\ref{alg:IBZ} is correct for computing an IBZ from the Brillouin zone and the group $\G$ of point operators of the space group of a crystal structure. The algorithm works even when the the atomic basis breaks some of the point symmetries of the lattice.

To begin we need a precise definition of an IBZ and the interior of a half-space or polytope.
\begin{definition}
For any half-space $H = \{x \in \R^n | d(x,v) \le d(x,v') \} \subset \R^n$, the \textit{interior} $\mathring{H}$ 
of $H$ is the set 
\[
\mathring{H} = \{x \in \R^n | d(x,v) < d(x,v') \}.
\] 
For any closed polytope $\Z\subset \R^n$ defined as the intersection $\Z = \bigcap_{H\in \C} H$ of a finite collection $\C$ of closed half-spaces, the \textit{interior} $\mathring{\Z}$ of $\Z$ is the set 
\[
\mathring{\Z} = \bigcap_{H\in \C} \mathring{H}.
\]  
\end{definition}

\begin{definition}\label{def:IBZ}
Given a Brillouin zone (BZ) consisting of a closed polytope $\P \subset \R^n$ with finite symmetry group $\G$, an \textit{irreducible Brillouin zone (IBZ)} is a closed polytope $\Q\subset \P$ such that
\begin{enumerate}
    \item\label{it:cover} For every point $x \in \P$, there exists a $g\in \G$ such that $g x \in Q$.
    \item\label{it:no-duplicate} For every point $y \in \mathring{\Q}$ in the interior of $\Q$ and every $g\in\G$, if $g y \neq y$, then $g y \not\in \Q$.
\end{enumerate}
\end{definition}

Note that in our definition an IBZ has no interior points that are equivalent under the action of $\G$, but it can have equivalent faces or edges.  Equivalent faces or edges may occur when the crystal structure has fewer point symmetries than the lattice, and they may come up even when the atomic basis does not  break symmetry.  To find an IBZ without equivalent faces or edges, any face of $\Q$ that is symmetrically equivalent to another face would need to be removed. See Fig. \ref{fig:symm-equiv} for an example in 2D.  For BZ integration, the fact that two faces are equivalent under the action of $\G$ poses no fundamental problem because the faces have measure zero (contribute nothing to the integral).

\begin{theorem}
Assume a finite symmetry group of a polytope $\P$ (the BZ) is $\G \subset \operatorname{O}(n)$ (every operator $g\in \G$ acts linearly and preserves distance in $\R^n$) and that $\P$ is the intersection of a collection $\M$ of half-spaces.  Assume further that the group $\G$ acts faithfully on the set $\V$ of vertices of $\P$, meaning that if $gv = v$ for all $v\in \V$, then $gx = x$ for all $x\in \P$.  Under these assumptions, Algorithm~\ref{alg:IBZ} correctly computes an IBZ as the intersection of the half-spaces returned by that algorithm. 
\end{theorem}

\begin{figure}[t]
\begin{center}
\includegraphics[width=0.6 \linewidth]{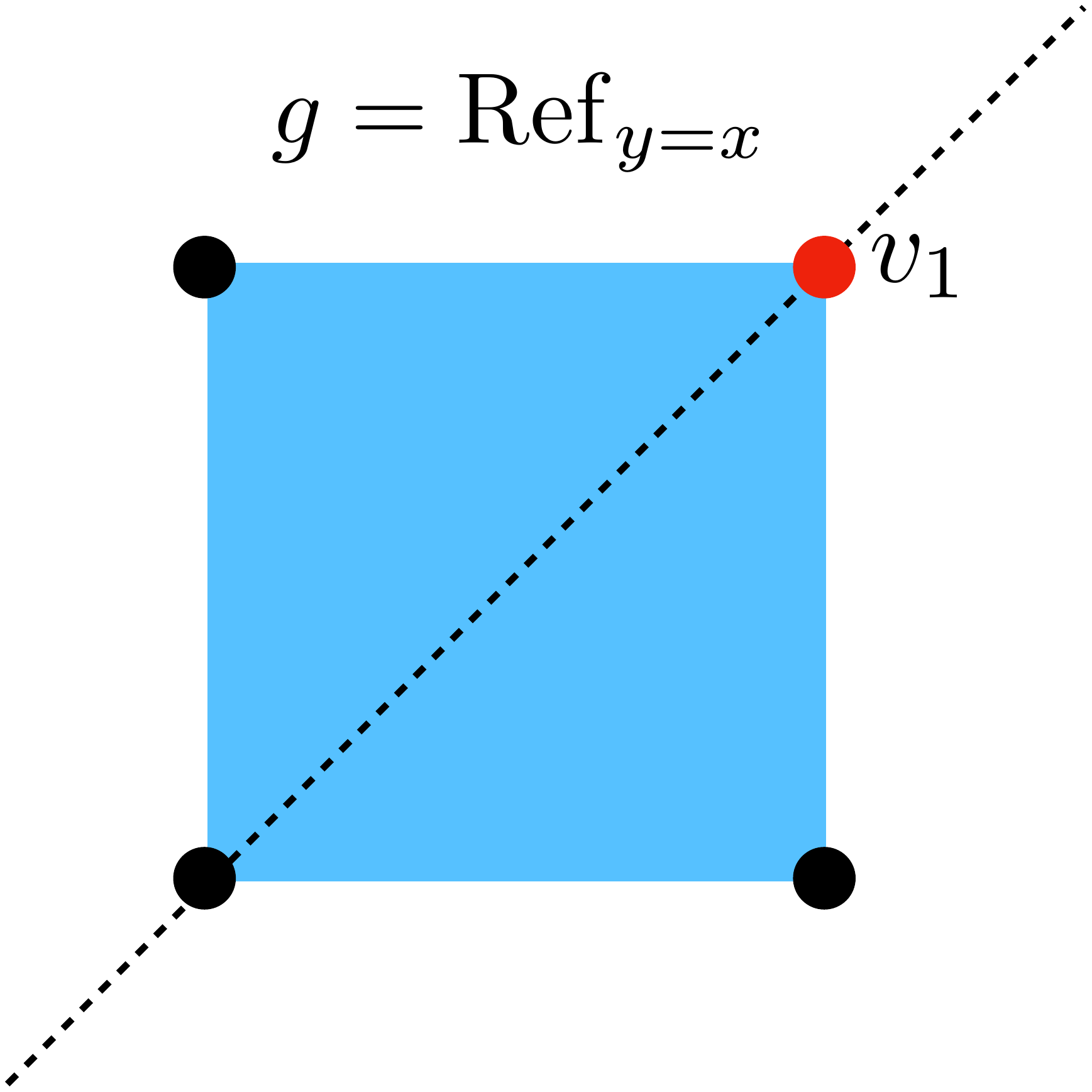}
\caption{The stabilizer $\stab_G(v_1)$ is the set of operators in $G$ that fix $v_1$.  Here, for the square lattice, the reflection $g = \operatorname{Ref}_{y=x}$ about the line $y=x$ leaves the point $v_1$ unmoved ($v_1 = g v_1$), so $g\in \stab_G(v_1)$.}
\label{fig:stabilizer}
\end{center}
\end{figure}

\begin{proof}   
Begin by giving $V = \{v_1,\dots, v_m\}$ the ordering used in the algorithm (in the loop starting at line 7).  Let $\G_0 = \G$, and for each $k\ge 1$ let
\[
\G_k = \stab_{\G_{k-1}} (v_k) = \{ g \in \G_{k-1} | g v_k = v_k\},
\] 
be the stabilizer, inside of $G_{k-1}$ of $v_k$; see figure~\ref{fig:stabilizer}. 
Let
\[
\N_k =  \{ H_{v_k, g}\}_{g\in G_{k-1} \setminus G_k}.
\]
Note that $\N_k$ is the set of all the hyperplanes associated to vertex $v_k$ that are added to $\N$ by the algorithm. 
Since $\G$ acts faithfully on $V$, only the identify element lies in the stabilizer of every vertex, so we have $F = \G\setminus \{I\} = \bigcup_{k=0}^{m-1} \G_k \setminus G_{k+1}$.

If $\N_0 = \M$ is the set of hyperplanes defining $\P$, then 
the final state of $\N$ is  
\[
\N = \bigcup_{k = 0}^m \N_k.
\]
Let $\P_0 = \P$, and for any $\ell \ge 1$ let
\[
\P_\ell = \bigcap_{k=0}^\ell \bigcap_{H \in \N_k} H
\] 
be the polytope constructed by intersecting all the hyperplanes added for all the vertices $v_1,\dots, v_\ell$.   We can characterize the polytope $\P_1$ as 
\begin{align*}
\P_1 
&= \{x \in \P_0 | d(x,v_1) \le d(x, g v_1)\, \forall g \in \G_0 \setminus \G_1\}\\
&= \{x \in \P_0 | d(x,v_1) \le d(x, g v_1)\, \forall g \in \G_0\}.
\end{align*}
And more generally, we can characterize the polytope $P_\ell$ as 
\begin{align}
\P_\ell 
&= \{x \in \P_{\ell-1} | d(x,v_\ell) \le d(x, g v_\ell)\, \forall g \in \G_{\ell-1} \setminus \G_\ell\} \notag\\
&= \{x \in \P_{\ell-1} | d(x,v_\ell) \le d(x, g v_\ell)\, \forall g \in \G_{\ell-1}\}.\label{eq:Pell}
\end{align}
The algorithm stops on or before vertex $v_m$ and returns $\N$, from which we can construct $\P_m$, which we show below satisfies the requirements to be an IBZ. 

To see that Condition~\ref{it:cover} of Definition~\ref{def:IBZ} holds, consider any $x_0 \in P_0 = P$.  For each $\ell \ge 1$ we will iteratively choose $g_\ell \in \G_{\ell-1}\setminus G_\ell$ such that $x_\ell= g_\ell x_{\ell-1}$ lies in $P_\ell$.  Therefore, we will have $g_m g_{m-1} \cdots g_2 g_1 x_0 \in \P_m$, as required.

For each $\ell \ge 1$, assume we are given $x_{\ell-1} \in \P_{\ell-1}$. Since $\G_{\ell-1}$ is finite, there exists $g \in \G_{\ell-1}$ that minimizes the distance $d(x_{\ell-1}, g v_\ell)$; that is, $d(x_{\ell-1}, g v_\ell) \le d(x_{\ell-1},h v_\ell)$ for any $h\in \G_{\ell-1}$.  Operating by $g^{-1}$ gives $d(g^{-1} x_{\ell-1}, v_\ell) \le d(g^{-1} x_{\ell-1}, g^{-1} h v_\ell)$ for all $h\in \G_{\ell-1}$.  But the set $\{ g^{-1}h | h\in \G_{\ell-1}\}$ is equal to the entire group $\G_{\ell-1}$.  Thus we have 
\begin{equation}\label{eq:gxell}
d(g^{-1}x_{\ell-1}, v_\ell) \le d(g^{-1} x_{\ell-1}, \gamma v_\ell)   \qquad \forall\gamma \in \G_{\ell-1}.
\end{equation}
Setting $g_{\ell} = g^{-1}$ and using Equations~\eqref{eq:Pell} and \eqref{eq:gxell} gives $g_\ell x_{\ell-1} \in \P_\ell$.
Iterating from $\ell=1$ to $\ell=m$  shows that $g_m g_{m-1} \cdots g_2 g_1 x_0 \in \P_m$, as required.  Thus Condition~\ref{it:cover} holds.

\begin{figure}[t]
\begin{center}
\includegraphics[width=0.5\linewidth]{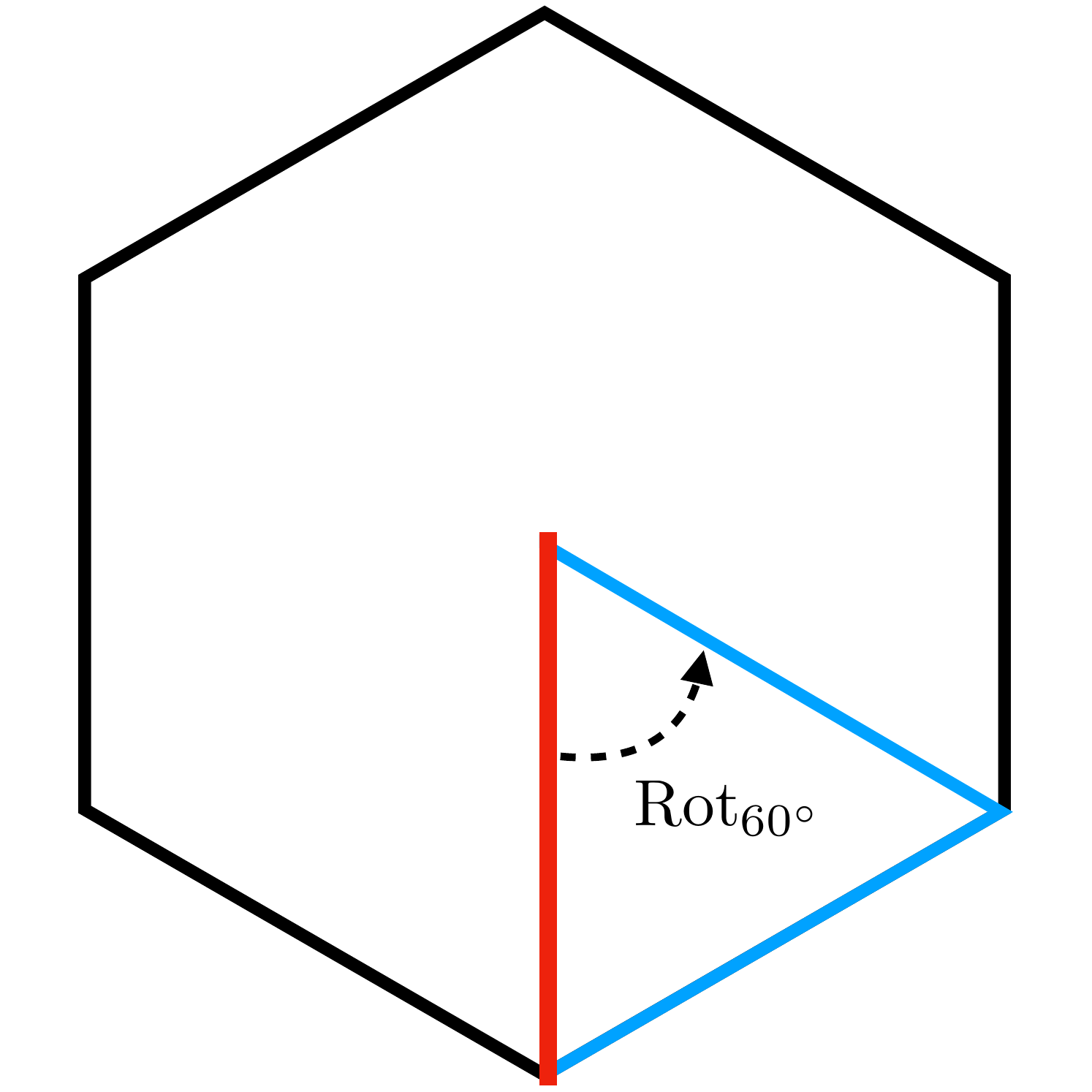}
\caption{Example of symmetrically equivalent IBZ edges. In the figure, the BZ is outlined in black and the IBZ in blue. In this example, the atomic basis has broken reflection symmetries; only rotational symmetries remain. The IBZ edge in red is symmetrically equivalent to the opposite edge by a $60^\circ$ rotation, which is an operation in the point group of a hexagonal lattice.}
\label{fig:symm-equiv}
\end{center}
\end{figure}

To see that Condition~\ref{it:no-duplicate} in Definition~\ref{def:IBZ} holds, first note that for every $v\in V$ if $g\in F$ satisfies $gv \neq v$, then we also have $g^{-1}v \neq v$.  This implies that if $H_{v,g}\in \N$, then $H_{v,g^{-1}}\in \N$.  
 Now consider any $y\in \mathring{\P}_m$ and any operator $g\in \G$ with $gy \neq y$.  Since $\G$ acts faithfully on $\V$, there exists at least one vertex $v\in \V$ such that $gv \neq v$.  Let $v$ be the first such $v \in V$ encountered in the loop (at line 7) over vertices in $\V$, so that $H_{v,g}$ and $H_{v,g^{-1}} \in \N$.  Since $y\in \mathring{\P}_m$, we must have $y\in \mathring{H}_{v,g^{-1}}$; hence
\begin{equation}\label{eq:last}
    d(gy,gv)  = d(y,v) < d(y, g^{-1}v) = d(gy, v), 
\end{equation}
where the two equalities in \eqref{eq:last} follow from the fact that any operator $g\in \G \subset O(n)$ preserves distances in $\R^n$.
This implies that $gy \not\in H_{v,g}$, and thus that $g y\not \in \P_m$.  Therefore Condition~\ref{it:no-duplicate} holds.
\end{proof}

\bibliography{ibz-writeup.bbl}
\end{document}